\documentclass[twocolumn,preprintnumbers,amsmath,amssymb, 14pt]{revtex4}
\usepackage{amsmath}
\usepackage{amssymb}
\usepackage{amsmath,amsthm,amssymb,amscd}
\usepackage{latexsym}
\usepackage{indentfirst}
\usepackage{subfigure}
\usepackage{graphicx}

\begin{document}

\title{
  \bf Comment on \lq\lq Length-dependent translation of messenger RNA by ribosomes"
}

\author{Yunxin Zhang}
\affiliation{Shanghai Key Laboratory for Contemporary Applied Mathematics, Laboratory of Mathematics for Nonlinear Science, Centre for Computational Systems Biology, and School of Mathematical Sciences, Fudan University, Shanghai 200433, China. {\rm Email: xyz@fudan.edu.cn}.
}

\baselineskip=16pt

\begin{abstract}
\normalsize\baselineskip=16pt
In the recent paper of Valleriani {\it et al} [Phys. Rev. E {\bf 83}, 042903 (2011)], a simple model for describing the translation of messenger RNA (mRNA) by ribosomes is presented, and an expression of the translational ratio $r$, defined as the ratio of translation rate $\omega_{\rm tl}$ of protein from mRNA to degradation rate $\omega_p$ of protein, is obtained. The key point to get this ratio $r$ is to get the translation rate $\omega_{\rm tl}$. In the study of Valleriani {\it et al}, $\omega_{\rm tl}$ is assumed to be the mean value of measured translation rate,
i.e. the mean value of ratio of the translation number of protein to the lifetime of mRNA. 
However, in experiments different methods might be used to get $\omega_{\rm tl}$. Therefore, for the sake of future application of their model to more experimental data analysis, in this comment three methods to get the translation rate $\omega_{\rm tl}$, and consequently the translational ratio $r$, are provided. Based on one of the methods which might be employed in most of the experiments, we find that the translational ratio $r$ decays exponentially with the length of mRNA in prokaryotic cells, and decays reciprocally with the length of mRNA in eukaryotic cells. This result is slight different from that obtained in Valleriani's study.
\end{abstract}


\maketitle

In recent paper \cite{Valleriani2011}, Valleriani {\it et al} presented a simple model to describe the length-dependent translation properties of messager RNA (mRNA).
In their model, the mRNA degradation process is assumed to be governed by rate $\omega_r$, i.e. the probability density of lifetime $t$ of an intact mRNA is
\begin{equation}
\phi_U=\omega_r\exp(-\omega_rt).
\end{equation}
Meanwhile, the rate of ribosome entering the coding region of mRNA is assumed to be $\omega_{\rm on}$, and the degradation rate of protein is denoted by $\omega_p$.

To discuss the mRNA length-dependent properties of the translation to protein by ribosomes, in \cite{Valleriani2011} the expression of translational ratio, defined as
\begin{equation}
r={\omega_{\rm tl}}/{\omega_{p}}, 
\end{equation}
is obtained for translations in both prokaryotic and eukaryotic cells. Where $\omega_{\rm tl}$ is the translation rate from mRNA to protein by ribosomes.
Since the protein degradation rate $\omega_p$ is independent of mRNA, the essential point to analyze the mRNA length-dependent properties of the translational ratio $r$, is to get the expression of translation rate $\omega_{\rm tl}$.
Experimentally, there might be three methods to get $\omega_{\rm tl}$:

\noindent{\bf (I)}: $\omega_{\rm tl}$ is obtained as the mean value of the measured translation rate $f(t)$ from mRNA to protein, i.e.,
\begin{equation}
\omega_{\rm tl}=\langle f(t)\rangle=\int_0^\infty f(t)\phi_U(t)dt,
\end{equation}
where $f(t)=N(t)/T(t)$, $N(t)$ is the mean number of proteins that the mRNA will synthesize if degradation occurs at time $t$, and $T(t)$ is the lifetime of a mRNA before completely degraded ($t$ is the lifetime of an intact mRNA).

\noindent{\bf (II)}: $\omega_{\rm tl}$ is obtained as the ratio of mean number $\langle N(t)\rangle$ of proteins translated from one mRNA to the mean lifetime $\langle T(t)\rangle$ of a mRNA, i.e.,
\begin{equation}
\omega_{\rm tl}=\left.{\langle N(t)\rangle}\right/{\langle T(t)\rangle}.
\end{equation}

\noindent{\bf (III)}: $\omega_{\rm tl}$ is obtained as the reciprocal of the mean duration time of translating one protein from mRNA, 
\begin{equation}
\omega_{\rm tl}=\frac{1}{\langle 1/f(t)\rangle}=\frac{1}{\langle T(t)/N(t)\rangle}.
\end{equation}

One can easily show that, for the mRNA translation problem discussed in \cite{Valleriani2011}, $T(t)=t$, $N(t)=\theta(t-t_L^{\rm pro}) \omega_{\rm on}(t-t_L^{\rm pro})$ for translation in prokaryotic cells, and $T(t)=t+t_L^{\rm eu}$, $N(t)=\omega_{\rm on}t$ for translation in eukaryotic cells. Where $\theta(t)$ is Heaviside function, i.e. $\theta(t)=1$ for $t>0$ and $\theta(t)=0$ for $t<0$, $t_L^{\rm pro}=L/v^{\rm pro}$ and $t_L^{\rm eu}=L/v^{\rm eu}$ are the time taken by ribosomes to reach the end of mRNA, $L$ is the length of mRNA, $v^{\rm pro}$ and $v^{\rm eu}$ are the average velocities of ribosome along mRNA in prokaryotic and eukaryotic cells respectively.

Intuitively, method {\bf (I)} is reasonable, and actually this method is used in \cite{Valleriani2011}. On the other hand, method {\bf (III)} is usually employed for some mathematical problems. For example, to get the mean translation rate $\omega$ of a process which includes two sub-processes with rate $\omega_1$ and $\omega_2$, one usually does the following calculations
\begin{equation}
\omega=\frac{1}{\langle T\rangle}=\frac{1}{\langle T_1\rangle+\langle T_2\rangle}
=\frac{1}{1/\omega_1+1/\omega_2}=\frac{\omega_1\omega_2}{\omega_1+\omega_2}.
\end{equation}
It should be pointed out that, if the method {\bf (III)} is employed to get $\omega_{\rm tl}$, the experimental samples with no protein synthesized, i.e. samples for $N(t)=0$, should be discarded to avoid the infinite waiting time cases. Correspondingly, in the theoretical calculation in Eq. (5), the average $\langle T(t)/N(t)\rangle$ should be done for only large enough time $t\ge t_{\rm lim}$ which satisfies $N(t)\ge 1$, and the probability density $\phi_U$ should be changed accordingly, $\hat\phi_U(t)=\phi_U(t)/\int_{t_{\rm lim}}^\infty\phi_U(t)dt=\omega_r\exp[-\omega_r(t-t_{\rm lim})]$ for $t\ge t_{\rm lim}$. In our numerical calculations, we use $t_{\rm lim}^{\rm pro}=t_L^{\rm pro}+1/\omega_{\rm on}$ for translation in prokaryotic cells, and $t_{\rm lim}^{\rm eu}=1/\omega_{\rm on}$ for translation in eukaryotic cells.

Meanwhile, the simple method {\bf (II)} is often used in the experimental data statistics. For this mRNA translation problem, one can easily get the following results by method {\bf (II)}. For mRNA translation in prokaryotic cells,
\begin{equation}
\begin{aligned}
\omega_{\rm tl}^{\rm pro}
=&\left.{\left\langle\theta(t-t_L^{\rm pro})\omega_{\rm on}(t-t_L^{\rm pro})\right\rangle}\right/{\langle t\rangle}\cr
=&\omega_{\rm on}\exp(-\omega_rt_L^{\rm pro}).
\end{aligned}
\end{equation}
So the translational ratio $r^{\rm pro}=\omega_{\rm tl}^{\rm pro}/\omega_p$ decays exponentially with the length $L^{\rm pro}$ of mRNA. Meanwhile, for the mRNA translation in eukaryotic cells,
\begin{equation}
\omega_{\rm tl}^{\rm eu}
=\left.{\langle\omega_{\rm on}t\rangle}\right/{\langle t+t_L^{\rm eu}\rangle}
=\left.{\omega_{\rm on}}\right/({1+\omega_{r}t_L^{\rm eu}}).
\end{equation}
So, roughly speaking, the translational ratio $r^{\rm eu}=\omega_{\rm tl}^{\rm eu}/\omega_p$ decays reciprocally with the length of mRNA. Note: similar as the above discussion about the calculation of translation rate $\omega_{\rm tl}$ in prokaryotic cells by method {\bf III}, if the samples with $N(t)=0$ are discarded in  experimental measurements, the methods {\bf I} and {\bf II} should also be changed correspondingly, with a modified probability density $\hat\phi_U(t)$ for $t\ge t_{\rm lim}$ only.

For the sake of comparison, the theoretical results of translational ratio $r$ of {\it E. coli} and {\it S. cerevisiae}, obtained by the three methods, are plotted in Fig. \ref{Fig1}, with model parameters listed in Tab. \ref{table1}. Since method {\bf I} is used by Valleriani {\it et al.} in \cite{Valleriani2011}, the curves in Fig. \ref{Fig1} (and the corresponding model parameters in Tab. \ref{table1}) are copied from \cite{Valleriani2011}. One can see that,  compared with the experimental data, these three methods all work reasonably well, but with slightly different model parameter (see Tab. \ref{table1}). The most reasonable parameters should be the ones which are obtained by the same method as that used in the experimental measurements.

In conclusion, when one tries to apply the model presented in \cite{Valleriani2011} to real experimental data, the method used to get the translation rate $\omega_{\rm tl}$ should be chosen properly to be consist with experiments.

\noindent{\bf Acknowledgments}
This work was supported by the Natural Science Foundation of Shanghai
(Grant 11ZR1403700).

\newpage

\begin{widetext}

\begin{table}
  \centering
\caption{Model parameters $\omega_{\rm on}/\omega_p$ and $\omega_r$ obtained by fitting the theoretical results to experimental data of {\it E. coli} and {\it S. cerevisiae} respectively (see Ref. \cite{Valleriani2011} and references therein for the detailed description of the experimental data).}
\begin{tabular}{l|cc|cc}
  \hline
  \hline
  &\multicolumn{2}{|c|}{{\it E. coli}}&\multicolumn{2}{c}{{\it S. cerevisiae}}\\
  \hline
   & $\omega_{\rm on}/\omega_p$ & $\omega_r$(min$^{-1}$) & $\omega_{\rm on}/\omega_p$ & $\omega_r$(min$^{-1}$) \\
   Valleriani {\it et al. } & 708.2 &0.2 & 7.69$\times 10^3$ & 0.2 \\
   method {\bf II} & 734 & 0.61 & 6.34$\times 10^3$ &0.22\\
   method {\bf III} & 841.1 & 0.15 & 6.62$\times 10^3$ &0.02\\
  \hline\hline
\end{tabular}
  \label{table1}
\end{table}

\newpage
\begin{figure}
  \includegraphics[width=250pt]{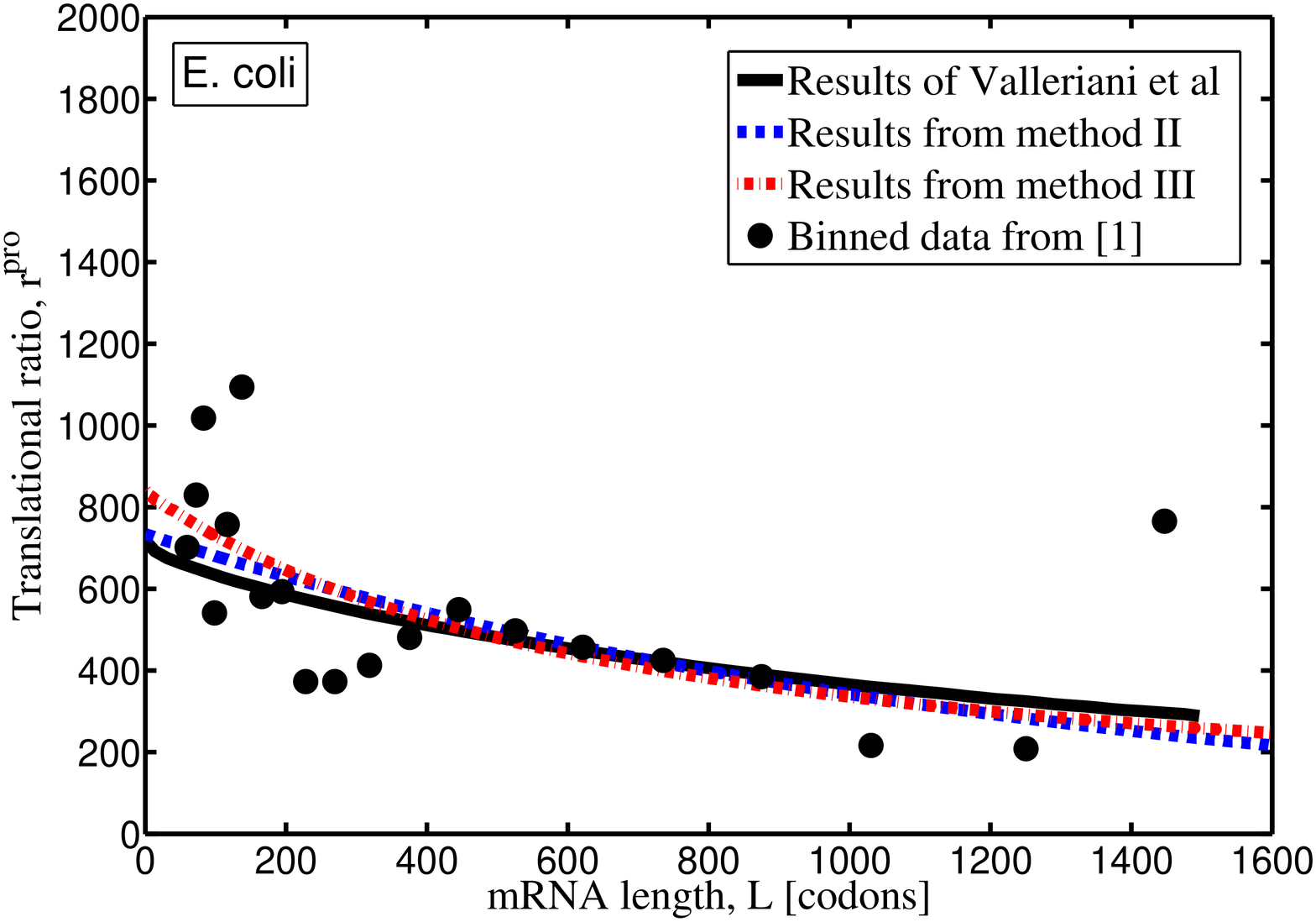}\includegraphics[width=250pt]{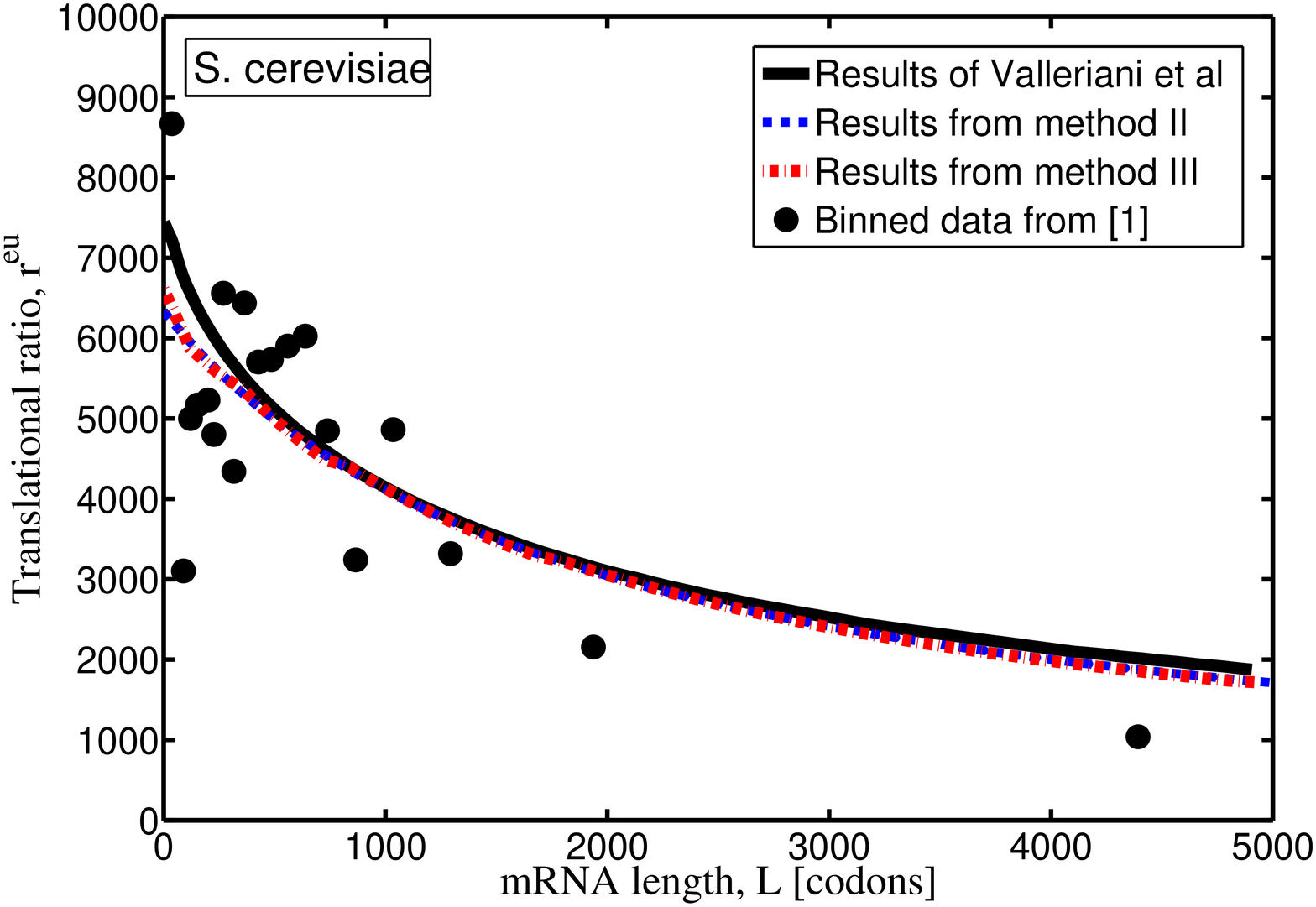}\\
  \caption{(Color online) Theoretical results of the translational ratio $r^{\rm pro}$ for {\it E. coli} (a) and $r^{\rm eu}$ for {\it S. cerevisiae} (b). The dots and solid curves are copied from Ref. \cite{Valleriani2011} (see the related references therein). The model parameters $\omega_{\rm on}/\omega_p$ and $\omega_r$ used in the calculations are listed in Tab. \ref{table1}.}\label{Fig1}
\end{figure}

\end{widetext}

\end{document}